**The Journal of Open Source Software**

# orbitize! v3: Orbit fitting for the High-contrast Imaging Community

**Sarah Blunt** [1,2,¶], **Jason Jinfei Wang** [1], **Lea Hirsch** [11], **Roberto Tejada** [12], **Vighnesh Nagpal** [13], **Tirth Dharmesh Surti** [5], **Sofia Covarrubias** [2], **Thea McKenna** [1,6,7,8], **Rodrigo Ferrer Chávez** [1], **Jorge Llop-Sayson** [4], **Mireya Arora** [14], **Amanda Chavez** [1], **Devin Cody** [9], **Saanika Choudhary** [1], **Adam J. R. W. Smith** [15], **William Balmer** [16], **Tomas Stolker** [3], **Hannah Gallamore** [15], **Clarissa R. Do Ó** [10], **Eric L. Nielsen** [15], and **Robert J. De Rosa** [17]

1 Center for Interdisciplinary Exploration and Research in Astrophysics (CIERA), Northwestern University 2 California Institute of Technology 3 Leiden Observatory, Leiden University 4 Jet Propulsion Laboratory, California Institute of Technology 5 Department of Physics/Kavli Institute for Particle Astrophysics and Cosmology, Stanford University 6 Southeastern Universities Research Association 7 Astronomy Department, Cornell University 8 NASA Goddard Space Flight Center, Code 698 9 Google LLC 10 Department of Astronomy & Astrophysics, University of California San Diego 11 Department of Chemical & Physical Sciences, University of Toronto Mississauga 12 Department of Astrophysical Sciences, Princeton University 13 University of California, Berkeley 14 University of Michigan 15 New Mexico State University 16 Johns Hopkins University 17 European Southern Observatory ¶ Corresponding author

## Summary

`orbitize!` is a package for Bayesian modeling of the orbital parameters of resolved binary objects from time series measurements. It was developed with the needs of the high-contrast imaging community in mind, and has since also become widely used in the binary star community. A generic `orbitize!` use case involves translating relative astrometric time series, optionally combined with radial velocity or astrometric time series, into a set of derived orbital posteriors.

This paper is published alongside the release of `orbitize!` version 3.0, which has seen significant enhancements in functionality and accessibility since the release of version 1.0 (Blunt et al., 2020).

## Statement of need

The orbital parameters of directly-imaged planets and binary stars can tell us about their present-day dynamics and formation histories (Bowler, 2016), as well as about their inherent physical characteristics (particularly mass, generally called "dynamical mass" when derived from orbital constraints, e.g. Brandt (2021), Lacour et al. (2021)).

`orbitize!` is widely used in the exoplanet imaging and binary star communities to translate astrometric data to information about eccentricities (Bowler et al., 2020), obliquities (Bryan et al., 2020), dynamical masses (Lacour et al., 2021), and more.

Each new released version of the `orbitize!` source code is automatically archived on Zenodo (Blunt et al., 2024).

## Major features added since v1

For a detailed overview of the `orbitize!` API, core functionality (including information about our Kepler solver), and initial verification, we refer the reader to Blunt et al. ([2020](#)). This section lists major new features that have been added to the code since the release of version 1.0 and directs the reader to more information about each of them. A full descriptive list of modifications to the code is maintained in our [changelog](#).

Major new features since v1 include:

1. The functionality to jointly fit the radial velocity (RV) time series for the primary star together with the secondary companion (see Section 3 of Blunt et al. ([2023](#))). For the primary star, the RV data can either be directly input into orbitize! (as explained in the [Radial Velocity Tutorial](#)) or be fitted separately and then used as priors (as detailed in the [Non-orbitize! Posteriors as Priors Tutorial](#)).

2. The ability to jointly fit absolute astrometry of the primary star. `orbitize!` can fit the Hipparcos-Gaia catalog of accelerations (Brandt ([2021](#)); see the [HGCA Tutorial](#)), as well as Hipparcos intermediate astrometric data and Gaia astrometry, following Nielsen et al. ([2020](#)) (see the [Hipparcos IAD Tutorial](#)). It can also handle arbitrary absolute astrometry (see the [Fitting Arbitrary Astrometry Tutorial](#)).

3. In addition to the MCMC and OFTI posterior computation algorithms documented in Blunt et al. ([2020](#)), orbitize! version 3 also implements a nested sampling backend, via dynesty (Speagle ([2020](#)); see the [dynesty Tutorial](#).)

4. `orbitize!` version 3 implements two prescriptions for handling multi-planet effects. Keplerian epicyclic motion of the primary star due to multiple orbiting bodies, following Lacour et al. ([2021](#)), is discussed in the [Multi-planet Tutorial](#), and N-body interactions are discussed in Covarrubias et al. ([2022](#)). The Keplerian epicyclic motion prescription only accounts for star-planet interactions, treating the motion of the star as a sum of Keplerian orbit signals, while the N-body prescription models this effect as well as planet-planet interactions.

5. The ability to fit in different orbital bases (Ferrer-Chávez et al. ([2021](#)), Surti et al. ([2023](#)); see the [Changing Bases](#) tutorial), as well as the ability to apply the observation-based priors derived in O'Neil et al. ([2019](#)) (see the [Observation-based Priors Tutorial](#)).

## Verification and Documentation

`orbitize!` implements a full stack of automated testing and documentation building practices. We use GitHub Actions to automatically run a suite of unit tests, maintained in [orbitize/tests](#), each time code is committed to the public repository or a pull request is opened. The Jupyter notebook tutorials, maintained in [orbitize/docs/tutorials](#), are also run automatically when a pull request to the `main` branch is opened. Documentation is built using `sphinx`, and hosted on readthedocs.org at [orbitize.info](#). We also maintain a set of longer-running tests in [orbitize/tests/end-to-end-tests](#) that show real scientific use cases of the code. These tests are not automatically run.

`orbitize!` is documented through API docstrings describing individual functions, which are accessible on [our readthedocs page](#), a set of [Jupyter notebook tutorials](#) walking the user through a particular application, a set of [frequently asked questions](#), and an in-progress ["manual"](#) describing orbit fitting with `orbitize!` from first principles.

## Comparison to Similar Packages

Since the release of `orbitize!` version 1, other open source packages have been released that have similar goals to `orbitize!`, notably orvara (Brandt et al., 2021) and octofitter (Thompson et al., 2023). This is a wonderful development, as all packages benefit from open sharing of knowledge! `orbitize!`, orvara, and octofitter can do many similar things, but each has unique features and strengths; as an example, octofitter is extraordinarily fast, and enables joint astrometry extraction and orbit modeling, while `orbitize!` has unique abilities to fit arbitrary absolute astrometry (i.e. not from Hipparcos or Gaia) and model data using an N-body backend. orvara analytically marginalizes over parallax assuming a prior informed by Gaia, a significant speed advantage, while `orbitize!` allows different parallax priors to be used. We recommend that users of each package compare the implementations of the particular features they wish to use.

Best practices for orbit fitting, particularly using radial velocities, for which the treatment of stellar activity is an active area of research, and absolute astrometry with Gaia and Hipparcos, for which the treatment of correlated errors is an active area of research, evolve quickly. The philosophy of `orbitize!` is to, as much as possible, implement multiple approaches to a problem, evidenced by our multiple implementations of radial velocity joint fitting and absolute astrometry joint fitting (described above). For detailed information about our particular implementations, we refer the reader to our documentation.

## Acknowledgements

Our team gratefully acknowledges support from the Heising-Simons Foundation. S.B. and J.J.W. are supported by NASA Grant 80NSSC23K0280. E.L.N. is supported by NASA Grants 21-ADAP21-0130 and 80NSSC21K0958.

This paper describes additions made to `orbitize!` between versions 1.0.0 and 3.0.0. We are extremely grateful to Isabel Angelo, Henry Ngo, James Graham, Logan Pearce, and Malena Rice, who contribtued code to version 1.0.0, and are included as authors in Blunt et al. (2020).